\documentclass{article}      % Specifies the document class

\font\tensl=cmsy10
\usepackage{amssymb}
\title{{\bf Common Space of Spin and Spacetime}\footnote{Published on Foundations of Physics Letters, Vol.18, No.3, 243-258 in June of 2005}}  % Declares the document's title.
\author{{\bf Wei Min Jin}\\ \\ {\it 2209 Carrington Ct., Lexington, KY 40513}\footnote{New Address: 6744 Highlands Creek Loop, Lakeland, FL 33813} \\{\it E-mail: weiminjin@netscape.net}\\{\it URL: http://www.weiminjin.com/}}      % Declares the author's name.
\date{Finalized on November 8, 2004}      % Deleting this command produces today's date.

\begin{document}             % End of preamble and beginning of text.

\maketitle                   % Produces the title.

\begin{center}
\begin{minipage}{135mm}
%\vskip0.2in
\begin{center} {} \end{center}
\baselineskip15pt
{Given Lorentz invariance in Minkowski spacetime, we investigate a common space of spin and spacetime. To obtain a finite spinor representation of the non-compact homogeneous Lorentz group including Lorentz boosts, we introduce an indefinite inner product space (IIPS) with a normalized positive probability. In this IIPS, the common momentum and common variable of a massive fermion turn out to be ``doubly strict plus-operators''. Due to this nice property, it is straightforward to show an uncertainty relation between fermion mass and proper time. Also in IIPS, the newly-defined Lagrangian operators are self-adjoint, and the fermion field equations are derivable from the Lagrangians. Finally, the nonlinear QED equations and Lagrangians are presented as an example.
}
\end{minipage}
\end{center}
%\vskip2in
\baselineskip15pt
\eject

\section* {1. INTRODUCTION}

Since the discovery of ``Dirac equation'' [Ref.1], there have been many years of development of relativistic quantum theories on Dirac fields. It is known that the Dirac equation is ``metrical'', in which spinor has more than one component. There are many references on spinors, like the book by Dirac [Ref.2] and the standard textbook by Bjorken and Drell [Ref.3]. The existence of spin in the Dirac equation, is a natural outcome of the unification of special relativity and quantum mechanics, and is one of the greatest triumphs of the Dirac theory. The relationship between spinor and spacetime has long been an interesting research subject, and there are many different approaches with a huge amount of publications along this line. But I have always been thinking about tackling these issues in a more straightforward and easy-to-follow kind of fashion. This paper is written exactly in this way. Hope it can bring some new results and insights to the development of Dirac quantum field theory.

Given those well-known results on spinors, I intend to explore the possibility of dealing with spin and spacetime in a simply unified fashion. The main idea is make use of a minimal complete set of variables with both spin and spacetime degrees of freedom, which I shall call ``common variable'' (see Jin [Ref.4]). It is then straightforward to construct a class of ``common operators'' such as ``common momentum'' out of this variable. I shall give this space, established directly on the idea of ``common variable'', the name ``common space'' for clarity. There are physical as well as mathematical reasons to construct such a ``common space''. Physically, spin, an ``intrinsic'' property of the electron or other spin-half fermions, appears naturally with respect to the ``external'' Lorentz transformations in Minkowski spacetime. There is always a sense of correspondence between spin and spacetime. Mathematically, this ``common space'' possesses a complete set of degrees of freedom in both spin and spacetime, and transforms in an ``invariant'' way, with spin and spacetime each transforming in a ``covariant'' way with respect to the Lorentz transformations. It is a simple union of spin and spacetime, in the sense of special relativity.

One of the most surprising results I come up with, is that the ``common momentum'' and ``common variable'' of a massive fermion, constructed in this ``common space'', are exactly the kind of linear self-adjoint ``doubly strict plus-operators'' in an indefinite inner product space, as sought-after and discussed in detail in an excellent mathematical book by Bognar [Ref.5]. There is a long history of research in indefinite inner product spaces, starting from Dirac [Ref.6], then Pauli [Re.7], Heisenberg [Refs.8,9] and many other researchers. The idea is to find a finite representation like the Dirac spinor representation for the homogeneous Lorentz group including non-compact Lorentz boosts, which can only be given in indefinite inner product space. There are some unresolved issues in this field, such as the definitions of positive probability, vacuum state, etc. I will provide my answers to these questions in this paper. I will also prove an uncertainty relation between fermion mass and proper time, by utilizing the commutation relation of ``common momentum'' and ``common variable''. Finally, I will present Dirac quantum field theory in the notion of ``common space'' with the introduction of so-called Lagrangian operators, which need to be self-adjoint in the indefinite inner product space to preserve Lorentz invariance. Nonlinear QED equations and Lagrangians [Ref.4] will be given as an example to demonstrate this approach.

\section* {2. COMMON SPACE}      % Produces section heading.  Lower-level
                             % sections are begun with similar 
                             % \subsection and \subsubsection commands.

To begin with, we define a common variable in terms of Dirac matrices $\gamma^\mu$ and spacetime coordinates:
$$\Omega = \gamma^\mu X_\mu,\eqno(2.1)$$
where $\gamma^\mu$ are expressed in the standard representation [Ref.3]. The derivative with respect to this variable can be deduced as
$$\partial_\omega=({1\over{\partial_\mu \Omega}})\partial_\mu=\gamma^\mu\partial_\mu. \eqno(2.2)$$
Hence a common momentum can be introduced as
$$P_\omega = i\partial_\omega=\gamma^\mu P_\mu.\eqno(2.3)$$ 
The natural units are to be used throughout, unless specified otherwise. It is straightforward to show the following commutation relation 
$$[\Omega, P_\omega] = -4i,\eqno(2.4)$$
by the anticommutation relations of $\gamma^\mu$: $\{\gamma^\mu,\gamma^\nu\}=2g^{\mu\nu}$, and the commutation relations of $X_\mu$ and $P_\nu$: $[X_\mu,P_\nu]=-ig_{\mu\nu}$, where $g^{\mu\nu}$ or $g_{\mu\nu}$ are spacetime metric elements with Minkowski signature $(1,-1,-1,-1)$. Here we have utilized the commutation relations of time and energy as well as space and momentum, discussed by Stuecklberg [Ref.10], Horwitz and Piron [Ref.11].

In common space, the Lorentz transformation $L_\omega$ is a direct product of a spinor one $L_s$ and a coordinate one $L_c$: $L_\omega=L_sL_c$. Since momentum $P_\mu$ are covariant in spacetime
$$L_cP_\mu L_c^{-1}=a_\mu^\nu P_\nu,\eqno(2.5)$$
and Dirac matrices $\gamma^\mu$ are contravariant in spinor space
$$L_s\gamma^\mu L_s^{-1}=a_\nu^\mu\gamma^\nu,\eqno(2.6)$$
it is easy to check that
$$L_\omega P_\omega L_\omega^{-1}=P_\omega,\eqno(2.7)$$
namely, common momentum is Lorentz invariant, and so are any other common operators to be introduced in common space.

Now for any two common operators 
$$Y_\omega = \gamma^\mu Y_\mu,\eqno(2.8a)$$
$$Z_\omega = \gamma^\mu Z_\mu,\eqno(2.8b)$$
we define a scalar product that eliminates spin components and provides a scalar in spacetime only:
$$Y_\omega\cdot Z_\omega = Y_\mu Z^\mu.\eqno(2.9)$$
A simple example is $P_\omega^2=P_\omega\cdot P_\omega=P_\mu P^\mu$.

Since the common momentum squared of a free particle with mass $m$ is constant
$P_\omega^2 = E^2 - {\bf p}^2 = m^2$, 
we have a wave equation of the second-order Klein-Gordon type
$$P_\omega^2\Psi(\Omega) = m^2\Psi(\Omega),\eqno(2.10)$$ 
and two wave equations of the first-order Dirac type
$$P_\omega\Psi(\Omega) = \pm m\Psi(\Omega).\eqno(2.11)$$
In this simple case, $P_\omega$ is a physical observable with eigenvalues $\pm m$.

The common momentum and common variable so-defined are not hermitian
but instead pseudohermitian: 
$$\gamma^0 P_\omega^\dagger\gamma^0 = P_\omega,\eqno(2.12a)$$
$$\gamma^0 \Omega^\dagger\gamma^0 = \Omega.\eqno(2.12b)$$
Consequently they are not well-defined in positive-definite Hilbert
space where all self-adjoint operators are hermitian. To
make them physical observables, we need to find a better way.

\section* {3. INDEFINITE INNER PRODUCT SPACE}

Back in 1942, Dirac [Ref.6] initiated the idea of implementing an indefinite inner product space. This idea was subsequently explored by Pauli [Ref.7] who found a self-adjointness requirement: the operators of physical observables in such a space be pseudohermitian rather than hermitian. Heisenberg [Refs.8,9] later tried to implement this idea in his unified field theory as well.

In a positive-definite Hilbert space, the invariance group is restricted to the compact unitary group including rotations and inversions, while the Lorentz boosts which are non-compact and non-unitary (though unimodular) are not included. This situation can be clearly stated according to Heisenberg [Ref.8]: ``a finite representation of a non-compact group can be given only in space with indefinite metric''.

Since the early work by these pioneers, there has been a lot of progress in this field. There is an excellent book by Bognar [Ref.5] on the mathematical aspects of indefinite inner product spaces. In this book, emphasis has been put on Krein spaces, the most important type of inner product spaces, which stand for non-degenerate, decomposable, and complete spaces. Interested readers should study it for detailed mathematical treatments. We shall only utilize some important definitions and results from this book, which are directly relevent to our approach. In the rest of this paper, we shall assume all our inner product spaces are Krein spaces.  

For our purpose, we define an indefinite inner product space (IIPS) for Dirac fields, which is an indefinite Krein space, with the following inner product
$$<\psi, \phi> = \int d\tau\overline\psi\phi,\eqno(3.1)$$
where $d\tau\equiv d^4x$, and $\overline\psi = \psi^\dagger\gamma^0$ is the adjoint of $\psi$. It can be divided into three subspaces with zero, positive and negative norms respectively
$$\hbox{\tensl Z}=\Bigl\{\psi|<\psi,\psi>\hskip0.05in=0\Bigr\},\eqno(3.2a)$$
$$\hbox{\tensl P}=\Bigl\{\psi|<\psi,\psi>\hskip0.05in>0\Bigr\},\eqno(3.2b)$$
$$\hbox{\tensl N}=\Bigl\{\psi|<\psi,\psi>\hskip0.05in<0\Bigr\}.\eqno(3.2c)$$
A simple example for a zero norm subspace is $\psi=cI\in\hbox{\tensl Z}$, a four-component column unit vector $I$ multiplied by a constant $c$. It is the negative norm subspace that causes trouble and confusion. 

In such an IIPS, the common momentum $P_\omega$ and common variable $\Omega$ are well-defined physical observables since they are both self-adjoint operators, namely, their expectation values are real
$$<P_\omega>={{\int d\tau\overline\psi P_\omega\psi}\over{\int
d\tau\overline\psi\psi}} = <P_\omega>^\dagger,\eqno(3.3a)$$
$$<\Omega>={{\int d\tau\overline\psi\Omega\psi}\over{\int
d\tau\overline\psi\psi}}=<\Omega>^\dagger.\eqno(3.3b)$$ 
To make the above integrals converge, we need to assume certain boundary conditions and let $\psi\not\in\hbox{\tensl Z}$. 

Mathematically, a symmetric operator in IIPS obeys: 
$$<\psi,O^*\phi>=<O\psi,\phi>=<\psi,O\phi>,\eqno(3.4)$$ 
namely its adjoint is 
$$O^*=\gamma^0 O^\dagger\gamma^0=O.\eqno(3.5)$$
A self-adjoint operator in IIPS is a symmetric operator with dense domain. We shall assume all our symmetric operators satisfying (3.5) have dense domain in IIPS, and simply call them self-adjoint operators. In this sense, common momentum and common variable are self-adjoint by (2.12a) and (2.12b). Note that if $O$ commutes with $\gamma^0$: $[O,\gamma^0]=0$, then it is hermitian. 

A transformation $L$ keeping the indefinite inner product invariant yields: 
$$<L\psi, L\phi> = <\psi, \phi>.\eqno(3.6)$$
This is true for almost all the homogeneous Lorentz transformations including non-compact Lorentz boosts [Ref.3], which satisfy
$$\gamma^0 L_\omega^\dagger\gamma^0 L_\omega = I.\eqno(3.7)$$
Note that if a Lorentz transformation commutes with $\gamma^0$: $[L_\omega,\gamma^0] = 0$, then it is a unitary transformation such as a rotation or a space inversion. It is special, however, for unitary time inversion $T_\omega$ [Ref.4] that satisfies
$$\gamma^0 T_\omega^\dagger\gamma^0 T_\omega = -I. \eqno(3.8)$$ 
Given that the time integral by $dt$ reverses sign under $T_\omega$, unitary time inversion also renders indefinite inner product invariant in (3.6). These results show that the so-defined indefinite inner product space provides a finite spinor representaion for the whole non-compact homogeneous Lorentz group.  

A major concern about the legitimacy of IIPS is whether one can come up with a valid interpretation of positive probability [Ref.9]. If simply defining a probability density as
$$\rho_0(x)=\overline\psi(x)\psi(x),\eqno(3.9)$$
one would get into trouble for $\rho_0(x)<0$ when $\psi(x)\in\hbox{\tensl N}$. To avoid this difficulty, we need to redefine the probability density in such a way that total probability adds up to one:
$$\rho(x)={{\overline\psi(x)\psi(x)}\over{<\psi,\psi>}}, \hskip0.1in \int\rho(x) d\tau = 1.\eqno(3.10)$$  
With this definition, we now have a normalized positive probability density $\rho(x)>0$ no matter $\psi(x)\in\hbox{\tensl P}$ or $\hbox{\tensl N}$ unless $\psi(x)\in\hbox{\tensl Z}$. This new definition is more natural, in the sense that it provides not only a generalization from nonrelativistic to relativistic quantum mechanics by preserving Lorentz invariance, but also a reasonable physical interpretation in a statistical fashion. 

In Chapter II of Ref.5, there are some detailed discussions about so-called fundamental symmetry $J$, which are quite analogous to the way we define our normalized positive probability. One can set $J=P^+-P^-$ with the fundamental projectors $P^+$ and $P^-$ onto the positive and negative subspaces respectively. Then one can define a $J$-inner product
$$(\psi,\chi)_J=(J\psi,\chi)=(\psi^+,\chi^+)-(\psi^-,\chi^-), \eqno(3.11)$$
which is positive-definite, namely, $(\psi,\psi)_J\geq0$.

\section* {4. DOUBLY STRICT PLUS-OPERATOR}

We need some definitions out of Bognar's book, for subsequent discussions. A linear operator in IIPS is called a ``plus-operator'', if it is defined everywhere in IIPS and carries non-negative vectors into non-negative ones. A ``plus-operator'' is called a ``strict plus-operator'', if it carries positive vectors into positive ones. A ``strict plus-operator'' is called a ``doubly strict plus-operator'', if its adjoint is also a ``strict plus-operator''. 

For a linear operator $T$ in IIPS, define 
$$\mu(T)={infimum}_{<\psi,\psi>=1}<T\psi,T\psi>.\eqno(4.1)$$ 
Then it turns out that $T$ is a plus-operator if $\mu(T)\geq0$; a strict plus-operator if $\mu(T)>0$; and a doubly strict plus-operator if $\mu(T)>0$ and $\mu(T^*)>0$ where $T^*$ is the adjoint of $T$.

For a massive fermion, it happens that its common momentum and
common variable are both ``time-like'', namely 
$$<P_\omega^2>={{<P_\omega\psi,P_\omega\psi>}\over{<\psi,\psi>}}>0,\eqno(4.2a)$$
$$<\Omega^2>={{<\Omega\psi,\Omega\psi>}\over{<\psi,\psi>}}>0,\eqno(4.2b)$$
therefore they are ``strict plus-operators''. Furthermore, they are self-adjoint, as mentioned in the last section, so their adjoints are also ``strict plus-operators''. We thus come to a very important conclusion: the common momentum and common variable of this massive fermion fall into the category of ``doubly strict plus-operators''. 

A nice feature about this kind of operators, denoted by $O$ now, is that the three subspaces are separable
$$O(\hbox{\tensl Z};\hbox{\tensl P};\hbox{\tensl N})\in\hbox{\tensl
Z};\hbox{\tensl P};\hbox{\tensl N}, \eqno(4.3)$$
and it is easy to see from (4.2a) and (4.2b) that
$<P_\omega\psi,P_\omega\psi>$ and $<\Omega\psi,\Omega\psi>$ have the same
sign as $<\psi,\psi>$. This can be used to prove a more general result (see
Lemma II.8.7 of Ref.5):
$$<O\psi, O\psi'> = \mu(O)<\psi,\psi'>\hskip0.1in (\psi,\psi'\in \hbox{IIPS})\eqno(4.4)$$ 
where $\mu(O) >0$ for doubly strict plus-operator $O$.

If $<\psi_0,\psi_0>=0$ in the zero norm subspace, then $<O\psi_0,O\psi_0>=0$ by (4.4). Let $\psi_1=O\psi_0$, then $<\psi_1,\psi_1>=0$ and $<{\psi_0+\psi_1},{\psi_0+\psi_1}>=0$ lead to $Re<\psi_0,\psi_1>=0$. Here $<\psi_0,O\psi_0>$ cannot be complex for $O$ is self-adjoint. It has to be zero. In any case, $<\psi_0,O\psi_0>=0$ whenever $<\psi_0,\psi_0>=0$. As a result, no physical observables can be measured in such a zero norm state. Heisenberg once called it ``ghost'' state [Ref.8]. Here we prefer to call it ``vacuum'' state. 

The conventional vacuum state is defined by 
$$P_\mu|\chi_0>=0.\eqno(4.5)$$
Let $\psi_0=\chi_0\otimes I$, we get 
$$P_\omega|\psi_0>=0,\eqno(4.6)$$
and $<P_\omega\psi_0,P_\omega\psi_0>=0$. From (4.4) we have $<\psi_0,\psi_0>=0$. Hence all $\psi_0$ which have zero norm become our new vacuum state in IIPS, related to the conventional vacuum states $\chi_0$ in the above sense.

\section* {5. UNCERTAINTY RELATION}

The simplest type of plus-operators can be expressed by a doubly strict plus-operator $O$ multiplied by a constant number $c$, since it is obvious (for $\psi\not\in\hbox{\tensl Z}$) 
$${{<cO\psi,cO\psi>}\over{<\psi,\psi>}}={{c^*c<O\psi,O\psi>}\over{<\psi,\psi>}}\geq0.\eqno(5.1)$$
The operators $cP_\omega$ and $c\Omega$ are both plus-operators though they may not be self-adjoint or doubly strict when constant $c$ is complex or simply zero. Furthermore any operator of the type $c_1P_\omega+c_2\Omega$ is a plus-operator by the superposition of indefinite inner product subspaces. This can be stated as follows: if $\psi\in\hbox{\tensl P}$, then $c_1P_\omega\psi \hskip0.05in(and\hskip0.05in c_2\Omega\psi)\in\hbox{\tensl P}+\hbox{\tensl Z}$, we have $(c_1P_\omega+c_2\Omega)\psi\in\hbox{\tensl P}+\hbox{\tensl Z}$; similarly if $\psi\in\hbox{\tensl N}$, we have $(c_1P_\omega+c_2\Omega)\psi\in\hbox{\tensl N}+\hbox{\tensl Z}$. In general, so long as $\psi\not\in\hbox{\tensl Z}$, we always have an inequality for any constant numbers $c_1$ and $c_2$:
$${{<(c_1P_\omega+c_2\Omega)\psi,(c_1P_\omega+c_2\Omega)\psi>}\over{<\psi,\psi>}}\geq0.\eqno(5.2)$$

Let us define a norm ratio as a function of a real parameter $\xi$ 
$$R(\xi)={{<(P_\omega-i\xi\Omega)\psi,(P_\omega-i\xi\Omega)\psi>}\over{<\psi,\psi>}}$$
$$=<P_\omega^2>+i\xi<[\Omega,P_\omega]>+\xi^2<\Omega^2>.\eqno(5.3)$$
To have nonzero physical observables in any non-vacuum state, from (5.2) we know $R(\xi)\geq0$ for any real
parameter $\xi$. That is to say 
$$<i[\Omega,P_\omega]>^2-4<P_\omega^2><\Omega^2>\leq0.\eqno(5.4)$$
The same argument can be applied to the fluctuations of common momentum and
common variable: 
$$\delta P_\omega=P_\omega-<P_\omega>,\eqno(5.5a)$$
$$\delta\Omega=\Omega-<\Omega>.\eqno(5.5b)$$
Noting that from (2.4) 
$$<[\delta\Omega,\delta P_\omega]>=<[\Omega,P_\omega]>=-4i,\eqno(5.6)$$
we have the following uncertainty inequality 
$$\Delta P_\omega\Delta\Omega\geq2\hbar,\eqno(5.7)$$
where 
$$\Delta P_\omega=\sqrt{<(\delta P_\omega)^2>},\eqno(5.8a)$$
$$\Delta\Omega=\sqrt{<(\delta\Omega)^2>}.\eqno(5.8b)$$
For a free particle $\Delta P_\omega=\Delta m$ and $\Delta\Omega=\Delta\tau$,
we have an uncertainty relation between its mass and proper time 
$$\Delta m\Delta\tau\geq2\hbar.\eqno(5.9)$$
So there exists uncertainty in measuring particle mass, given
the uncertainty in locating its spacetime position. This uncertainty relation was also discussed early by Arshansky and Horwitz [Ref.12].

There is an alternative way to prove inequality (5.4) by utilizing Schwarz
inequality in a semi-definite inner product space (see Lemma I.2.2 of Ref.5): 
$$(x,x)(y,y)\geq|(x,y)|^2.\eqno(5.10)$$
Now we define a positive-definite inner product 
$$(Q\psi,Q'\psi)={{<Q\psi,Q'\psi>}\over{<\psi,\psi>}},\eqno(5.11)$$
for any plus-operators $Q$ and $Q'$ with $\psi\not\in\hbox{\tensl Z}$. We know $P_\omega$ and $i\Omega$ are both strict plus-operators, namely, 
$$(P_\omega\psi,P_\omega\psi)>0,\eqno(5.12a)$$
$$(i\Omega\psi,i\Omega\psi)>0.\eqno(5.12b)$$
From the Schwarz inequality (5.10) we get 
$$(P_\omega\psi,P_\omega\psi)(i\Omega\psi,i\Omega\psi)\geq|(P_\omega\psi,i\Omega\psi)|^2$$
$$\geq|Re(P_\omega\psi,i\Omega\psi)|^2={1\over4}|(\psi,i[\Omega,P_\omega]\psi)|^2.\eqno(5.13)$$
This is the same as (5.4). It is similar to prove (5.7) by the Schwarz inequality.

\section* {6. QUANTUM FIELD THEORY}

In quantum field theory, one normally starts from a Lagrangian, then applies variational principles to derive field equations. Using our notations, we write an action as an integral of a Lagrangian density {\tensl L}
$$W=\int_R d\omega\hbox{\tensl L}(\psi,\partial_\omega\psi),\eqno(6.1)$$
where $d\omega\equiv \sqrt{(d\Omega)^2} = d^4x$ is a measure in spacetime, and $R$ is a four-dimensional region with a three-dimensional boundary $B$ that can be either fixed or not fixed (see Barut [Ref.13]). Here we discuss fixed boundary only. 

To render the whole approach manifestly Lorentz invariant, we introduce a Lagrangian operator
$\hbox{\^{\tensl L}}$ in the indefinite inner product space, so that the action can be written as 
$$W=\int_R d\omega\overline\psi(x)\hbox{\^{\tensl
L}}\psi(x)\equiv<\psi,\hbox{\^{\tensl L}}\psi>.\eqno(6.2)$$
Since the action is real-valued, the Lagrangian operator ought to be self-adjoint, namely, 
$$\hbox{\^{\tensl L}}^*=\gamma^0\hbox{\^{\tensl
L}}^\dagger\gamma^0=\hbox{\^{\tensl L}}.\eqno(6.3)$$
This is a stringent condition in our approach, in contrast to the conventional approach of choosing Lagrangians with much arbitrariness.

The variation of the action (6.1) with respect to field variables is given by
$$\delta W=\int_R d\omega({{\partial\hbox{\tensl L}}\over{\partial\psi}}\delta\psi+{{\partial\hbox{\tensl L}}\over{\partial\partial_\omega\psi}}\delta\partial_\omega\psi)$$
$$=\int_R d\omega[({{\partial\hbox{\tensl
L}}\over{\partial\psi}}-{{\partial\hbox{\tensl
L}}\over{\partial\partial_\omega\psi}}\partial_\omega^\leftarrow)\delta\psi+\partial_\omega({{\partial\hbox{\tensl
L}}\over{\partial\partial_\omega\psi}}\delta\psi)].\eqno(6.4)$$
In the first term, the common operator $\partial_\omega^\leftarrow$ indicates that the derivatives are acting
on the left while the order of matrix multiplication is not altered. In
the second term, we define 
$$\partial_\omega({{\partial\hbox{\tensl
L}}\over{\partial\partial_\omega\psi}}\delta\psi)={{\partial\hbox{\tensl
L}}\over{\partial\partial_\omega\psi}}\partial_\omega^\leftarrow(\delta\psi)+{{\partial\hbox{\tensl
L}}\over{\partial\partial_\omega\psi}}\partial_\omega(\delta\psi).\eqno(6.5)$$
Its integral can be changed to a surface integral by Gauss' theorem
$$\int_R
d\omega[\partial_\omega({{\partial\hbox{\tensl L}}\over{\partial\partial_\omega\psi}}\delta\psi)]=\int_B
d\sigma_\omega\cdot\hbox{\tensl F}_\omega,\eqno(6.6)$$
where the surface common operator is defined by 
$$\sigma_\omega=\gamma^\mu\sigma_\mu,\eqno(6.7)$$
with $\sigma_\mu$ being a surface four-vector, and 
$$\hbox{\tensl F}_\omega={{\partial\hbox{\tensl L}}\over{\partial\partial_\omega\psi}}\delta\psi.\eqno(6.8)$$
Considering matrix multiplication, we rewrite the right-hand side of (6.6)
$$\int_B d\sigma_\omega\cdot\hbox{\tensl F}_\omega=\int_B ({{\partial\hbox{\tensl L}}\over{\partial\partial_\omega\psi}})d\sigma_\omega(\delta\psi).\eqno(6.9)$$
 
Then let $\delta W=0$ and field vanish on the surface, we get the following equation of motion 
$${{\partial\hbox{\tensl L}}\over{\partial\psi}}-({{\partial\hbox{\tensl L}}\over{\partial\partial_\omega\psi}})\partial_\omega^\leftarrow=0.\eqno(6.10)$$
The same argument is true when $\overline\psi$ is the field variable, which
leads to 
$${{\partial\hbox{\tensl L}}\over{\partial\overline\psi}}-\partial_\omega({{\partial\hbox{\tensl L}}\over{\partial(\overline\psi\partial_\omega^\leftarrow)}})=0.\eqno(6.11)$$

For a simple free particle, we write a Lagrangian density
$$\hbox{\tensl L}=\overline\psi i\partial_\omega\psi-m\overline\psi\psi.\eqno(6.12)$$
From (6.11), we get the free Dirac equation 
$$i\partial_\omega\psi-m\psi=0.\eqno(6.13)$$
In general, for any c-number potential $V(\psi,\overline\psi)$ not involving the derivatives of field variables, we introduce a self-adjoint Lagrangian operator satisfying (6.3) 
$$\hbox{\^{\tensl L}}=i\partial_\omega-m-V(\psi,\overline\psi),\eqno(6.14)$$
and write a Lagrangian density
$$\hbox{\tensl L}=\overline\psi i\partial_\omega\psi-m\overline\psi\psi-\overline\psi V(\psi,\overline\psi)\psi.\eqno(6.15)$$ 
From (6.11), the equation of motion turns out to be
$$i\partial_\omega\psi-m\psi-V\psi-\overline\psi{{\partial V}\over{\partial\overline\psi}}\psi=0.\eqno(6.16)$$ 

In the case of electrodynamical interaction, we identify 
$$V(\psi,\overline\psi)={1\over2}e\overline\psi\gamma^\mu\psi A_\mu,\eqno(6.17)$$
to obtain a nonlinear equation of motion [Ref.4]: 
$$i\partial_\omega\psi-m\psi-eJ^\mu A_\mu\psi=0.\eqno(6.18)$$
Here four-current $J^\mu=\overline\psi\gamma^\mu\psi$ satisfies a continuity equation 
$$\partial_\mu J^\mu=0,\eqno(6.19)$$
and four-potential $A_\mu$ is generated by an external four-current 
$$\square A_\mu=qJ_\mu^{ext},\eqno(6.20)$$
under Lorentz gauge condition 
$$\partial^\mu A_\mu=0.\eqno(6.21)$$
Equations (6.18)-(6.21) are what we need to establish a nonlinear QED in which Lorentz gauge is unique and inherent [Ref.4]. 

Defining a common current and a common potential
$$J_\omega=\gamma^\mu J_\mu,\eqno(6.22a)$$
$$A_\omega=\gamma^\mu A_\mu,\eqno(6.22b)$$
we rewrite the above set of equations by utilizing scalar product (2.9)
$$i\partial_\omega\psi-m\psi-e(J_\omega\cdot A_\omega)\psi=0,\eqno(6.23)$$
$$\partial_\omega\cdot J_\omega=0,\eqno(6.24)$$
$$\Box A_\omega=qJ_\omega^{ext},\eqno(6.25)$$
$$\partial_\omega\cdot A_\omega=0,\eqno(6.26)$$
where $\Box=\partial_\omega\cdot\partial_\omega=\partial_\mu\partial^\mu$. 

This theory is also invariant under the translations in the inhomogeneous Lorentz group (or Poincare group), with common momentum as the generator. Given a state vector
$\psi(\Omega_0)$ at a particular common variable $\Omega_0$, similar to the Schr\"odinger picture 
$$i\partial_\omega\psi(\Omega_0)=P_\omega\psi(\Omega_0),\eqno(6.27)$$
we have a state vector at another common variable $\Omega+\Omega_0$
$$\psi(\Omega+\Omega_0)=[\exp(-iP_\omega\cdot\Omega)]\psi(\Omega_0).\eqno(6.28)$$
We may also have a definition in analogy to the Heisenberg picture. Given a physical common operator $O_\omega(\Omega_0)$ at $\Omega_0$, we define this operator at another common variable $\Omega+\Omega_0$ by 
$$O_\omega(\Omega+\Omega_0)=[\exp(iP_\omega\cdot\Omega)]O_\omega(\Omega_0)[\exp(-iP_\omega\cdot\Omega)].\eqno(6.29)$$
Then we have an equation of motion 
$$\partial_\omega\cdot O_\omega=i[P_\omega,O_\omega].\eqno(6.30)$$
This implies that common operator $O_\omega$ is conserved with respect to an evolution of $\Omega$, if it commutes with common momentum $P_\omega$.

In the case of electrodynamical interaction, comparing (6.23) with (6.27) we obtain its common momentum
$$P_\omega=m+eJ_\omega\cdot A_\omega,\eqno(6.31)$$
where mass and charge are physical observables as part of common momentum. This provides an alternative way to unravel the mystery of mass and charge. It is easy to verify
$$[P_\omega,J_\omega]=0,\eqno(6.32)$$
to obtain conservation equation (6.24), namely continuity equation (6.19). 

We now define a conjugate common momentum of field $\psi$
$$\Pi={{\partial\hbox{\tensl L}}\over{\partial\partial_\omega\psi}}.\eqno(6.33)$$
By the action principle, a generator that provides conservation laws in integral form, can be defined as follows
$$\int_\sigma d\sigma_\omega\cdot\hbox{\tensl F}_\omega=\int_\sigma \Pi(d\sigma_\omega)\delta\psi,\eqno(6.34)$$
with $\sigma$ being a three-dimensional surface. For electrodynamical interaction (6.17), Lagrangian (6.15) gives a conjugate common momentum
$$\Pi=i\overline\psi,\eqno(6.35)$$
and a generator
$$\int_\sigma d\sigma_\omega\cdot\hbox{\tensl F}_\omega=\int_\sigma
d\sigma_\mu(i\overline\psi\gamma_\mu\delta\psi),\eqno(6.36)$$
leading to the continuity equation (6.19) in differential form as well.

For any arbitrary separation, the anticommutation relations of Dirac field operators can be written as   
$$\Bigl\{\psi(\Omega),\psi(\Omega')\Bigr\}=0,\eqno(6.37a)$$
$$\Bigl\{\overline\psi(\Omega),\overline\psi(\Omega')\Bigr\}=0,\eqno(6.37b)$$
$$\Bigl\{\psi(\Omega),\overline\psi(\Omega')\Bigr\}=-iS(\Omega-\Omega'),\eqno(6.37c)$$
with
$$S(\Omega)={1\over(2\pi)^4}\int dp_\omega{1\over{P_\omega+m}}\exp(iP_\omega\cdot\Omega),\eqno(6.38)$$
where $dp_\omega\equiv \sqrt{(dP_\omega)^2} = d^4p$ is a measure in energy-momentum space. The above anticommutation relations of canonical quantization always hold true, so long as the interaction potential in the Lagrangian (6.15) does not involve the derivatives of field variables. Note that there are no derivatives in our nonlinear QED interaction potential (6.17). 

\section* {7. REMARKS}

This paper presents a new way to develop relativistic quantum theory on Dirac fields. The emphasis has been put on the construction of a ``common space'' which is actually a simple union of ``intrinsic'' spin and ``external'' spacetime. A ``common variable'' in common space is a minimal complete set of variables in spin and spacetime. A ``common momentum'' in common space is defined in terms of the derivative with respect to ``common variable''. Along this line, a whole class of ``common operators'' can be defined in terms of Dirac matrices and spacetime operators, which are manifestly Lorentz invariant. These common operators are not hermitian but instead pseudohermitian, and are well-defined physical observables in an indefinite inner product space. 

In nonrelativistic quantum mechanics, the positive-definite Hilbert space has been widely used in Euclidean three-space with time being a parameter. While in relativistic quantum mechanics, spacetime is Minkowskian with time being an independent coordinate. The invariance group of 4-dimensional Minkowski spacetime is the non-compact Lorentz group including Lorentz boosts. For such a non-compact invariance group, the finite representation like the Dirac spinor representation discussed exclusively here in this paper, can only be given in indefinite inner product space. To resolve one of the major contradictions in this approach, we have defined a normalized positive probability, which is Lorentz invariant as well.

This paper furthur demonstrates that common momentum and common variable of a massive fermion are linear self-adjoint ``doubly strict plus-operators'' in the indefinite inner product space defined for Dirac fields. This critical result leads to the introduction of ``vacuum state'' in the zero norm subspace; to the uncertainty relation between fermion mass and proper time; and to the better understanding of Lorentz invariant physical quantities such as mass and charge. Recasting Dirac quantum field theory in the notion of common space, we have introduced self-adjoint Lagrangian operators in order to write the full action integral in the indefinite inner product space. By variational principles, we can then deduce quantum fermion field equations for different interactions, such as nonlinear quantum electrodynamical interaction. 

We may just as well generalize what we have done here into curved spacetime. First we may define an indefinite inner product space for Dirac fields with a generalized measure. Then we may define so-called common operators of spin and spacetime, such as common momentum by utilizing spin affine connections, and show that these common operators are symmetric or self-adjoint in the indefinite inner product space. In any static spacetime with time-independent metric, which includes Minkowski spacetime as a special case, we can utilize on-mass-shell projections to project fermion Fock space onto the positive and negative energy subspaces, and quantize free Dirac field operators in terms of the creation and annihilation operators in both positive and negative energy subspaces, as demonstrated in another paper by Jin [Ref.14]. 

For interacting Dirac fields, quantization procedures become too complex, if not impossible. Once more complicated interactions are involved, nonlinear effects tend to be dominant, and particle creation and annihilation are not really clear-cut during the processes. Many unstable intermediate particles appear in high energy physics experiments. Certain exotic states of matter may form at short length scale as discussed by C.P. Kouropoulos [Ref.15]. We eventually have to go back to some nonlinear field equations and try to find better ways to solve them in a nonperturbative but physically and mathematically attainable fashion. This is the kind of challenge we are still facing in solving nonlinear quantum field theories like nonlinear QED [Ref.4].

\section* {ACKNOWLEDGMENTS}

The authur is very grateful to a referee for many valuable comments.

\section* {REFERENCES}

1. P. A. M. Dirac, {\it Proc. Roy. Soc.} {\bf A117}, 610 (1928); {\bf A118}, 351 (1928)\\
2. P. A. M. Dirac, {\it Spinors in Hilbert Space} (Plenum, New York, 1974)\\
3. J. D. Bjorken and S. D. Drell, {\it Relativistic Quantum Mechanics} (McGraw-Hill, New York, 1964)\\
4. W.M. Jin, {\it Found. of Phys.} {\bf 30}, 1943 (2000); quant-ph/0001029\\
5. J. Bognar, {\it Indefinite Inner Product Spaces} (Springer-Verlag, Berlin, 1974)\\ 
6. P. A. M. Dirac, {\it Proc. Roy. Soc.} {\bf A180}, 1 (1942)\\
7. W. Pauli, {\it Rev. Mod. Phys.} {\bf 15}, 175 (1943)\\
8. W. Heisenberg, {\it Introduction to the Unified Field Theory of Elementary Particles} (John Wiley \& Sons, London, 1966)\\
9. W. Heisenberg, ``Indefinite metric in state space'' in {\it Aspects of Quantum Theory} (Salam and Wigner, eds.) (Cambridge, 1972)\\
10. E.C.G. Stucklberg, {\it Helv. Phys. Acta} {\bf 14}, 322, 589 (1941); {\bf 15}, 23 (1942)\\
11. L.P. Horwitz and C. Piron, {\it Helv. Phys. Acta} {\bf 46}, 316 (1973)\\
12. R. Arshansky and L.P. Horwitz, {\it Found. of Phys.} {\bf 15}, 701 (1985)\\
13. A.O. Barut, {\it Electrodynamics and Classical Theory of Fields and Particles} (Dover, New York, 1980)\\
14. W.M. Jin, {\it Class. Quantum Grav.} {\bf 17}, 2949 (2000); gr-qc/0009010\\
15. C.P. Kouropoulos, {\it Classically Bound Electrons - EVs, Exotic Chemistry $\&$ ``Cold Electricity''}, http://www.intalek.com/Index/Projects/Research/ColdElectricity.pdf 

\end{document}